\documentclass[twoside]{report}
\usepackage{iwsm}
\usepackage{graphicx}
\usepackage{booktabs}
\usepackage{amsmath, amssymb}
\usepackage{amsfonts}
\usepackage[framemethod=tikz]{mdframed}
\usepackage{algpseudocode}
\usepackage{algorithmicx}
\usepackage{algorithm2e}
\usepackage{mathtools}
\usepackage[colorlinks]{hyperref}
\hypersetup{
  colorlinks=true,
  linkcolor=red,
  citecolor=red
}
\AtBeginDocument{%
  \hypersetup{%
    linkcolor=.,%
    citecolor=blue,%
  }%
}%

\begin{document}


\title{Regularisation of Generalised Linear Mixed Models with autoregressive random effect}
\titlerunning{Regularised EM algorithms for GLMM with AR(1) random effect}

\author{Jocelyn Chauvet\inst{1}, 
Catherine Trottier\inst{1}\inst{2}, 
Xavier Bry\inst{1}}
\authorrunning{Chauvet et al.}    

\institute{Institut Montpelli\'erain Alexander Grothendieck, CNRS, Univ. Montpellier, France.
\and Univ. Paul-Val\'ery Montpellier 3, F34000, Montpellier, France. }

\email{jocelyn.chauvet@umontpellier.fr}

\abstract{
We address regularised versions of the Expectation-Maximisation (EM) algorithm for Generalised Linear Mixed Models (GLMM) in the context of panel data (measured on several individuals at different time-points). A random response $y$ is modelled by a GLMM, using a set $X$ of explanatory variables and two random effects. The first one introduces the dependence within individuals on which data is repeatedly collected while the second one embodies the serially correlated time-specific effect shared by all the individuals. Variables in $X$ are assumed many and redundant, so that regression demands regularisation. 
In this context, we first propose a $L_2$-penalised EM algorithm, 
and then a supervised component-based regularised EM algorithm 
as an alternative.
}


\maketitle


\section{Introduction}
One of the main purposes of panel data analysis is to account for the dependence induced by repeatedly measuring an outcome on each individual over time. 
Besides, due to the fact that it is nowadays increasingly possible to collect large amounts of data, the potentially high level of correlation among explanatory variables should be taken into account. 
To this end, ridge-, lasso- and component-based regularisations have recently been highlighted.  

In the Linear Mixed Models (LMM) framework, 
\textcolor{blue}{\hyperlink{Eliot}{Eliot et al. (2011)}} 
proposed to extend the classical ridge regression to longitudinal biomarker data. 
They suggested a variant of the EM algorithm to maximise a ridge-penalised likelihood. This variant includes a new step to  find the best shrinkage parameter --- in the Generalised Cross-Validation (GCV) sense --- at each iteration.

With a view towards variable selection, 
\textcolor{blue}{\hyperlink{Schelldorfer}{Schelldorfer et al. (2014)}} 
proposed a $L_1$-penalised algorithm for fitting a high-dimensional Generalised Linear Mixed Models (GLMM), using Laplace approximation and an efficient coordinate gradient descent.

In the GLM framework, in order both to regularise the linear predictor and to facilitate its interpretation, 
\textcolor{blue}{\hyperlink{Bry1}{Bry et al. (2013)}} 
developed a PLS-type method --- Supervised Component-based Generalised Linear Regression (SCGLR) --- which yields explanatory components. 
\textcolor{blue}{\hyperlink{Chauvet}{Chauvet et al. (2016) }}
extended SCGLR to GLMM by using an adaptation of Schall's algorithm 
(\textcolor{blue}{\hyperlink{Schall}{Schall (1991)}}).

To the best of our knowledge, the random effects in the previous strategies are assumed normally distributed with independent levels. 
However, in the panel data framework, the question naturally arises of the autocorrelation of the time-specific random effect. 
Consequently, our objective is twofold: on the one hand, to extend the Mixed Ridge Regression of \textcolor{blue}{\hyperlink{Eliot}{Eliot et al. (2011)}}  
to the GLMMs with an AR(1) random effect; and on the other hand, to present the main ideas of a new version of SCGLR which handles the high dimensional case.



\section{Model hypotheses}
In this section, we recall the main hypotheses of the GLMM framework and we introduce the random effect distributions.
For the sake of simplicity, we consider balanced panel data with $N$ individuals, each of them observed at the same $T$ time-points. We denote by $n = N \times T$ the total number of observations.
Let $X$ be the $n \times p$ fixed effects design matrix, and $U$ the $n \times q$ random effects design matrix. 
Let also $Y$ be the $n$-dimensional random response vector, $\beta$ the $p$-dimensional vector of fixed effects, and $\xi$ the $q$-dimensional vector of random effects. We observe a realisation $y$ of $Y$, but $\xi$ is not observed. We conventionally assume that:
\begin{itemize}
\item[(i)]
the $Y_i \, | \, \xi, \; i \in \left\lbrace 1, \ldots, n \right\rbrace$ are independent and their distribution 
belongs to the exponential family; 
\item[(ii)]
the conditional mean $\mu_i = \mathbb{E}(Y_i \, | \, \xi)$ 
depends on $\beta$ and $\xi$  
through \linebreak the link function $g$ and the linear predictor 
$\eta_i = x_i^\T\beta + u_i^\T \xi$, with 
$\eta_i = g(\mu_i)$.
\end{itemize}
Less conventionally, we consider two random effects $\xi_1$ and $\xi_2$ with different roles and distributions: 
\begin{itemize}
\item[(i)]
$\xi_1$ is the individual-specific random effect. Assuming individuals are independent, we suppose:
$$
\xi_1 \sim \mathcal{N}_N \left( 0, \, \sigma_1^2 I_N \right),
$$
with $\sigma_1^2$ the unknown ``individual'' variance component.
\item[(ii)]
$\xi_2$ is the serially correlated time-specific effect common to all the individuals, which can be viewed as some latent phenomenon not measured in the explanatory variables. As these effects tend to persist over time, we model them with a stationary order 1 autoregressive process (AR(1)), i.e. for each $t \in \left\lbrace 
1, \ldots, T-1 \right\rbrace$, 
\begin{align*}
\xi_{2,t+1} &= \rho \xi_{2,t} + \nu_t, \\ 
\nu_t &\overset{\text{iid}}{\sim} 
\mathcal{N} \left( 0, \, \sigma_2^2 \right),
\end{align*}
where $\rho$ is the unknown parameter of the AR(1) and $\sigma_2^2$ the unknown ``temporal'' variance component.
Such time-specific effects arise naturally for instance
in an economic context (e.g. all companies share a common economic climate which tend to persist over time), 
or in biology (e.g. the ecological environment is often too complex to be directly observed through the explanatory variables).   
\end{itemize}
Finally, $\xi_1$ and $\xi_2$ are assumed independent. 
Denoting 
${\xi=\left( \xi_1^\T, \xi_2^\T \right)^\T}$, \linebreak
${U_1=I_N \otimes \textbf{1}_T}$,
${U_2=\textbf{1}_N \otimes I_T}$ and  
${U=\left[ U_1 \, | U_2 \right]}$, 
linear predictor $\eta$ can be matricially written:
$$
\eta = X\beta + U\xi.
$$
\section{Methods}
Owing to the GLMM dependence structure, the Fisher scoring algorithm was adapted by 
\textcolor{blue}{\hyperlink{Schall}{Schall (1991)}}. 
We, in turn, adapt Schall's algorithm by introducing a regularised EM at each step in order to take into account the high level of correlation in $X$ and the unconventional random effects distributions. Two steps appear in our method: the linearisation step and the estimation step.

\medskip

{\bf \underline{Linearisation step}.}
For each $i \in \left\lbrace 1, \ldots, n \right\rbrace$, a 
classic order 1 linearisation of $y_i$ around $\mu_i$ is given by:
$g(y_i) \simeq z_i = g(\mu_i) + (y_i - \mu_i) g'(\mu_i)$.
Matricially, this approximation provides a working variable $z$ entering the following linearised model 
$$  \mathcal{M} : \quad
z = X\beta + U\xi + e,
$$
with $\text{Var}(e \,|\, \xi) = \text{Diag} 
\left( 
\left[ g'(\mu_i) \right]^2 \text{Var}(Y_i \,|\, \xi)
\right)_{i=1, \ldots, n} = \Gamma$.


\medskip

{\bf \underline{Estimation step}.}
Instead of solving Henderson's system associated with $\mathcal{M}$ seen as a LMM (as proposed by 
\textcolor{blue}{\hyperlink{Schall}{Schall (1991))}}, 
we rather propose a regularised EM step.
We suggest an adaptation of the $L_2$-penalised EM algorithm of 
\textcolor{blue}{\hyperlink{Eliot}{Eliot et al. (2011)}} 
for low dimensional data ($p<n$), 
and a supervised component-based regularised EM algorithm for the high dimensional case ($p \gg n$), because then, interpretable dimension reduction is needed.

\subsection{The low dimensional case}
Our estimation step is based on 
\textcolor{blue}{\hyperlink{Green}{Green (1990)}}, who popularised the use of the EM algorithm for penalised likelihood estimation, and 
\textcolor{blue}{\hyperlink{Golub}{Golub et al. (1979)}}, 
who encouraged the use of the GCV for efficiently choosing the ridge parameter 
$\lambda$. 
However, contrary to the homoskedastic LMM considered in 
\textcolor{blue}{\hyperlink{Eliot}{Eliot et al. (2011)}},
$\mathcal{M}$ contains heteroskedastic errors. 
We will then opt for the modified GCV criterion suggested by 
\textcolor{blue}{\hyperlink{Andrews}{Andrews (1991), p. 372}}.

Denoting $\theta = \left( \beta, \sigma_1^2, \sigma_2^2, \rho \right)$, 
\textcolor{red}{\autoref{Chauvet::PSEUDOCODE}} describes the current iteration of our $L_2$-penalised EM algorithm for GLMM with AR(1) random effect.

\medskip

\noindent
\fbox{
\begin{minipage}{\linewidth}
\begin{algorithm}[H]
\begin{itemize}
\item[\bf (1)] {\bf \underline{Linearisation step}.}
Set:
\vspace{-0.7em}
$$
\mathcal{M}^{[t]} : z^{[t]} = X\beta + U\xi + e, \; \text{with} \,  
\text{Var}(e \,|\, \xi) = \Gamma^{[t]}.  
$$
\item[\bf (2)] {\bf \underline{Estimation step}.}
\begin{itemize}
\vspace{-0.5em}
\item[\bf (2.a)]
Denoting $L$ the complete log-likelihood of the linearised model, 
define the associated complete penalised log-likelihood $L_{\text{pen}}$ by:
\vspace{-0.7em}
$$
L_{\text{pen}}(\theta ; z,\xi) = L(\theta ; z,\xi) - \frac{\lambda}{2} \beta^\T \beta.
$$ 
\vspace{-1em}
\item[\bf (2.b)]
Denoting $\widehat{z}^{[t]}$ the fitted values and $S_{\lambda}^{[t]}$ the ``hat-matrix'' satisfying the equality $\widehat{z}^{[t]} = S_{\lambda}^{[t]} z^{[t]}$, set:
\vspace{-0.7em}
$$
\lambda^{[t]} \longleftarrow \text{arg } \underset{\lambda}{\min} \left\lbrace
\text{GCV}(\lambda) = 
\dfrac{n^{-1} \left\lVert z^{[t]} - S_{\lambda}^{[t]} z^{[t]} 
\right\rVert^2_{ {\Gamma^{[t]}}^{-1}} }
{\left[  1 - n^{-1} \text{tr} \left( S_{\lambda}^{[t]} \right)  \right]^2}
\right\rbrace.
$$
\vspace{-0.7em}
\item[\bf (2.c)] \textbf{EM step.} Set:
\vspace{-0.7em}
\begin{align*}
\mathcal{Q}_{\text{pen}} \left(\theta, \theta^{[t]} \right) &= 
\mathbb{E}_{\xi|z}
\left[
L_{\text{pen}}(\theta ; z^{[t]},\xi) \,|\, \theta^{[t]}, \lambda^{[t]}
\right], \\
\theta^{[t+1]} &\leftarrow \text{arg } \underset{\theta}{\max} \;
\mathcal{Q}_{\text{pen}} \left(\theta, \theta^{[t]} \right).
\end{align*}
\end{itemize}
\item[\bf (3)] {\bf \underline{Updating step}.}
Set $\xi^{[t+1]} = \mathbb{E}_{\xi|z} \left( \xi \,|\, \theta^{[t+1]} \right)$,
and update working variable $z^{[t+1]}$ and variance-covariance matrix $\Gamma^{[t+1]}$.
\end{itemize}
\vspace{-0.2em}
Steps {\bf (1)-–(3)} are repeated until stability of parameters $\beta$, $\sigma_1^2$, $\sigma_2^2$ and $\rho$ is reached.
\bigskip
\caption{Current iteration of the $L_2$-penalised EM algorithm for GLMM with AR(1) random effect.}
\label{Chauvet::PSEUDOCODE}
\end{algorithm}
\end{minipage}
}

\subsection{The high dimensional case}
In the $p \gg n$ case, 
we need to decompose the linear predictor on a small number of interpretable dimensions. To that end, we propose
to iteratively maximise a component-based regularised $\mathcal{Q}-$function. 

Let $C~=~XU$ be the set of principal components of $X$ with non-zero eigenvalues and $f~=~Cw$ the component we currently seek.
Let also $\phi$ denote a structural relevance (SR) criterion
(see \textcolor{blue}{\hyperlink{Bry2}{Bry and Verron (2015)}}): 
\vspace{-0.5em}
$$
\phi(w) = \left(
\sum_{j=1}^p \left[ \text{cor}^2 \left( x^j, f \right) \right]^l
\right)^{\frac{1}{l}}, \quad l \geqslant 1.
$$
$s \in \left[ 0,1 \right]$ being a parameter tuning the relative importance of the SR with respect to $L$,  
the $\mathcal{Q}-$function would then be:
\vspace{-0.4em}
\begin{align*}
\mathcal{Q}_{\text{reg}} \left(\theta, \theta^{[t]} \right) &= 
\mathbb{E}_{\xi|z}
\left[
L_{\text{reg}}(\theta ; z,\xi) \,|\, \theta^{[t]}
\right], \, \text{with} \\
L_{\text{reg}}(\theta ; z,\xi) &= (1-s) L(\theta ; z,\xi) + s \phi(w).
\end{align*}
Parameters $s$ and $l$ are tuned by cross-validation and higher rank components are computed like rank 1 component, after adding extra orthogonality constraints. 

\section{Numerical results}
\begin{figure}[ht!]
\begin{center}
\includegraphics[width=\linewidth]{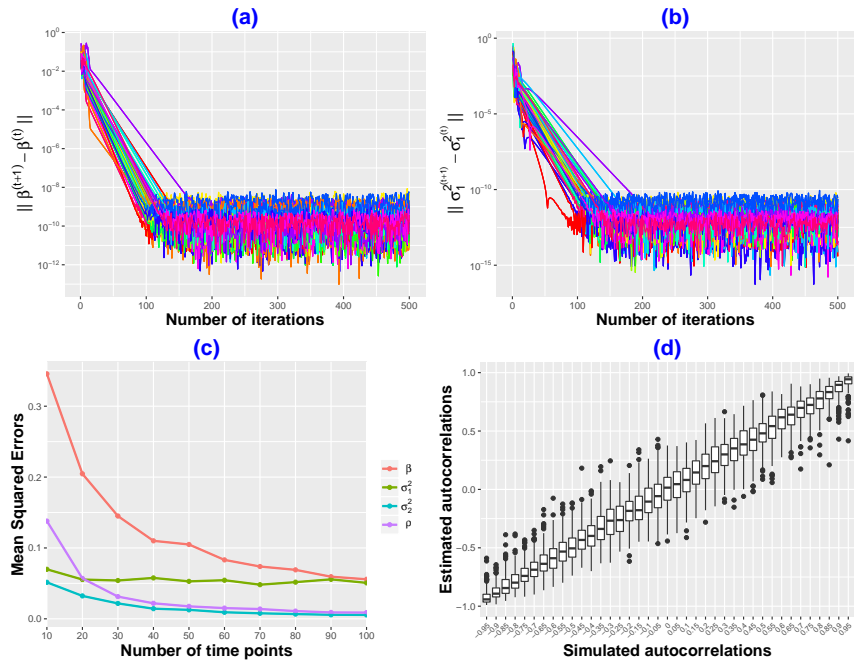}
\end{center}
\caption{
Graphical diagnoses relative to the $L_2$-penalised EM algorithm. 
\textbf{(a),(b):} 
\textit{40 trajectories of the $L_2$-convergence criterion 
for parameters $\beta$ and $\sigma_1^2$ 
(A similar behaviour is observed for parameters $\sigma_2^2$ and $\rho$). About a hundred iterations is necessary to achieve convergence.}
\textbf{(c):} 
\textit{MSEs of parameters $\beta, \sigma_1^2, \sigma_2^2$ and $\rho$ on simulated data where $N=10$ and 
$T \in \left\lbrace 10,20, \ldots, 100 \right\rbrace$. As expected, MSEs of 
$\beta, \sigma_2^2$ and $\rho$ decrease towards zero. In contrast, since $N$ is fixed, the MSE of $\sigma_1^2$ is constant.}
\textbf{(d):} 
\textit{Boxplots of estimated $\rho$ according to real value.} 
}
\label{Chauvet:diagnosticsgraphiques}
\end{figure}

In order to evaluate the performance of our $L_2$-penalised EM algorithm,
we conducted simulation studies in the canonical Poisson case.
We present some graphical diagnoses in 
\textcolor{red}{\autoref{Chauvet:diagnosticsgraphiques}}, which
aim at answering three questions: 
(1) Is the convergence assured? 
(2) How good are the estimations? 
(3) Are they sensitive to the value of $\rho$?
The answers to these questions is given in the figure's caption.

\references
\begin{description}
\hypertarget{Andrews}{
\item[Andrews, D.W.] (1991). 
        Asymptotic optimality of generalized CL, cross-validation, and
        generalized cross-validation in regression with heteroskedastic
        errors. 
        {\it Journal of Econometrics}, 
        {\bf 47}, 359\,--\,377. }
\vspace{-0.5em}        
\hypertarget{Bry1}{       
\item[Bry, X., Trottier, C., Verron, T., and Mortier, F.] (2013).
Supervised component generalized linear regression using a pls-extension of the fisher scoring algorithm. 
{\it Journal of Multivariate Analysis}, 
{\bf 119}, 47\,--\,60.}
\vspace{-0.5em}       
\hypertarget{Bry2}{
\item[Bry, X. and Verron, T.] (2015). 
THEME: THEmatic model exploration \linebreak through multiple co-structure maximization. 
{\it Journal of Chemometrics}, {\bf 29}, 637\,--\,647.}  
\vspace{-0.5em}             
\hypertarget{Chauvet}{
\item[Chauvet, J., Trottier, C., Bry, X., and Mortier, F.] (2016). $\,$
       Extension $\,$ to $\,$ mixed models of the Supervised Component-based
       Generalised Linear Regression. 
       {\it COMPSTAT: Proceedings in Computational Statistics}.} 
\vspace{-0.5em}    
\hypertarget{Eliot}{
\item[Eliot, M., Ferguson, J., Reilly, M.P., and Foulkes, A.S.] (2011).
       Ridge Regression for Longitudinal Biomarker Data.
       {\it The International Journal of Biostatistics}, 
       {\bf 7}, 1, Article 37.}
\vspace{-0.5em}
\hypertarget{Golub}{
\item[Golub, G.H., Heath, M., and Wahba, G.] (1979). 
       Generalized cross- validation  as a method for choosing a good ridge parameter. 
       {\it Technometrics}, 
       {\bf21}, 215\,--\,223.}
\vspace{-0.5em}
\hypertarget{Green}{
\item[Green, P.J.] (1990). 
       On use of the EM for penalized likelihood estimation. 
       {\it Journal of the Royal Statistical Society, Series B}, 
       {\bf 52}, 443\,--\,452.}
\vspace{-0.5em}
\hypertarget{Schall}{
\item[Schall, R.] (1991). 
       Estimation in generalized linear models with random effects. 
       {\it Biometrika}, 
       {\bf 78}, 719\,--\,727.}
\vspace{-0.5em}
\hypertarget{Schelldorfer}{
\item[Schelldorfer, J., Meier, L., and B\"{u}hlmann, P.] (2014). 
       Glmmlasso: an algorithm for high-dimensional generalized linear mixed models using $l_1$-  
       penalization.    
       {\it Journal of Computational and Graphical Statistics}, 
       {\bf 23}, 460\,--\,477.}
\end{description}

\end{document}